\documentclass[showpacs,aps,prd,eqsecnum,floats]{revtex4}
\newif\ifpdf\ifx\pdfoutput\undefined\pdffalse\else\pdfoutput=1\pdftrue\fi
\usepackage{graphicx}
\usepackage{amsfonts}
\usepackage{subfigure}
\usepackage{dcolumn}
\usepackage{bm}

\begin{document}

\title{\bf{POST-WMAP ASSESSMENT OF INFRARED CUTOFF IN THE PRIMORDIAL SPECTRUM FROM INFLATION}}

\author{Rita Sinha}
\email{rsinha@iucaa.ernet.in}

\author{Tarun Souradeep}
\email{tarun@iucaa.ernet.in}
\affiliation{Inter-University Center for Astronomy and Astrophysics,  
Post Bag 4, Ganeshkhind Pune 411007 India.}

\date{\today}

\begin{abstract}
The recent Cosmic Microwave Background (CMB) measurements indicate that there is power deficiency of the CMB anisotropies at large scales compared with the $\Lambda$CDM model. We have investigated the possibility of explaining such effects by a class of primordial power spectra which have infrared cutoffs close to the horizon scale. The primordial power spectrum recovered by direct deconvolution of the observed CMB angular spectrum indicates that the data prefers a sharp infrared cutoff with a localized excess (bump) just above the cutoff. We have been motivated to assess plausible extensions of simplest inflationary scenarios which readily accommodate similar form of infrared cutoff.  We carry out a complete Bayesian analysis of the parameter space using {\it Markov Chain Monte Carlo} technique with such a class of primordial power spectra. We show that primordial power spectrum that have features such as an infrared cutoff followed
by a subsequent excess in power give better fit to the observed data compared to a nearly scale-invariant power law or power spectrum with just a monotonic infrared cutoff. However, there is substantial room for improvement in the match to data and calls for exploration of other mechanisms that may lead to infrared cutoff even closer to that recovered by direct deconvolution approach.
\end{abstract}


\maketitle

\section{\label{sec:level1}Introduction\protect\\}

The Cosmic Microwave Background (CMB) is the most copious of all the isotropic backgrounds observed from radio to gamma rays. It has become the most important observational tool both for determining the global cosmological parameters and to understand the physics of the early universe. The concordant model of cosmology that has emerged from recent observations is the inflationary $\Lambda$CDM model. This describes the universe as spatially flat with primordial fluctuations which are adiabatic and Gaussian. The $\Lambda$CDM model has been quite successful in explaining many aspects of the universe conforming most with the observational data~\cite{{wmap_cp},{wmap_inflat},{bond_postmap},{boom05}}. The primordial power spectrum formed during the inflationary epoch, in the simplest case can be described as a single inflaton field with a nearly scale-invariant Harrison-Zel'dovich spectrum expressed as $\mathcal{P}(k)\propto k^{n_s-1}$, `$k$' being the comoving wavenumber. However, inflationary scenarios readily accommodate significant departures from scale invariance~\cite{{koflinde85},{koflinde87},{silkturn87},{kofpogo88},{bondsalo89},{mukhzelni},{polarstaro92}}. Distortions in the shape of the initial power spectrum imprints characteristic
scales on the form of $\mathcal{P}(k)$. The amount of such distortions manifested on the angular power spectrum and the scales at which they occur are strictly related to the nature of the primordial fluctuations and to the matter content of the universe. Careful analysis of all features of the CMB power spectrum put constraints on the entire parameter space. Recently a lot of attention has been devoted to the power anomalies at low multipoles~\cite{fnote1} compared to that obtained using the $\Lambda$CDM model of the universe. This was first reported by the Cosmic Background Explorer  (COBE)~\cite{cobe94} satellite and now
further confirmed by Wilkinson Microwave Anisotropy Probe (WMAP) observations~\cite{{wmap_res},{wmap_inflat},{wmap3y_hinshaw06}}. In the recent past many speculations have been made to provide a satisfactory explanation to this discrepancy but they have been successful only to a certain
extent.  The WMAP team have crudely fit the power at low scales by putting a running spectral index to the nearly scale-invariant power law~\cite{wmap_res}.  This aspect has also been projected as a probable new feature in the CMB anisotropies such as related to curvature scale~\cite{efst03}. This has been related to the topology of the universe as well, such as a dodecahedral universe~\cite{nature_03} or possibly a violation of statistical isotropy~\cite{amir0304}. A low quadrupole moment has been suggested to be originating also from dark energy, quintessence models with
isocurvature perturbations~\cite{{moroi04},{gord_Hu04}} via correlations with adiabatic density fluctuations which lower the
adiabatic Integrated Sachs-Wolfe contributions to the quadrupole. Even model-independent searches have been made to look for features in the CMB power spectrum~\cite{{bridle03},{hanne04},{pia03},{arman04},{pia05}}. The studies where the power spectrum has been reconstructed in a set of bins in k-space have the obvious disadvantage that the reconstruction depends on the width and the placement of the bins. This is also true for methods such as linear interpolation, wavelet, top-hat bins which can contribute to smoothening out or omission of features. Despite the progress the issue of low scale power still remains a subject of debate. The most crucial point perhaps is the limited information available at large angular scales due to cosmic variance~\cite{fnote2}. Although large cosmic variance at the quadrupole does make the power discrepancy less significant, it is important to bear in mind that cosmic variance is proportional to the `true' angular power spectrum of the underlying model. The cosmic variance is small for a model that predicts the observed low quadrupole.
In principle, the case for an infrared cutoff is best addressed by bayesian evidence study. However, different bayesian evidence studies for the models with infrared cutoff has yielded different conclusions regarding the evidence for an infrared cutoff in the primordial spectrum~\cite{jaf04,brid05}. This is possibly due to the dependence of bayesian evidence analysis on the different class of models explored, details of parameterization of models and the datasets used. Hence these results may not readily translate into a general conclusion on the evidence for `infrared cutoff' though ~\cite{jaf04} find that models with cutoff in power generally produce higher evidence compared to scale invariant or a single spectral index model.

We revisit the possibility of explaining this observed suppression of
power at low multipoles using known mechanisms of modifying the shape
of the primordial power spectrum from inflation to introduce an infrared
cutoff~\cite{{contaldi03_supcmb},{cline03},{db_inflat03},{dble_inflat04}}.
The limited success of model spectra with monotonic infrared cutoff
studied is apparent from the non-monotonic infrared cutoff indicated
by direct deconvolution of the primordial spectrum from the CMB
angular power spectrum using WMAP data~\cite{arman04,bump04}. The direct
approach suggests that the data favours a sharp infrared cutoff
followed by a localized excess (`bump'). As mentioned in
Ref.~\cite{arman04}, this is not uncommon since mechanisms that
produce a sharp change in amplitude (such as required for the infrared
cutoff) can also have an overshoot followed by damped
oscillations~\cite{fnote3}. Here, we focus on inflationary models with an infrared
cutoff followed by a localized excess power (referred to as `bump')
in the power spectrum as a feasible possibility in explaining this
deviation. This seems to be a promising avenue (such as
~\cite{BSI98}). Such features have also been seen by large-scale
structure
surveys~\cite{{cobe94},{einasto97},{gaz_baugh97},{bump04}}. We have
carried out an exhaustive parameter estimation with such a class of
primordial spectra using the Bayesian approach by applying the
{\it{Markov Chain Monte Carlo}} (MCMC) technique~\cite{cosmo}.

In the sections to follow we first describe briefly in section II, the
various forms of primordial power spectra considered in this
study. This is followed in section III by a brief overview of the
methodology adopted (MCMC). In Section IV we give the parameter
estimates with such a class of primordial power spectra which is
followed by the final section on discussion and conclusion.

\section{Primordial power spectra with features}  
While reconstructing the initial conditions of the universe it is
important to devise a method that can efficiently constrain the shape
of the power spectrum of primordial matter density fluctuations. It
has been shown in ~\cite{arman04} that the direct recovery of the
primordial power spectrum from the CMB (WMAP) observations indicate a
sharp infrared cutoff around the horizon scale ($k_* \sim k_h=
2\pi/\eta_0$, where $\eta_0$ is the present horizon) followed by a
`bump' ~\cite{fnote5_irl}. The resulting angular spectrum gives a significantly better
likelihood than a simple (scale-invariant) Harrison-Zel'dovich,
scale-free power spectrum and also a simple monotonic infrared cutoff.
This motivates us to consider a broader variety of primordial power
spectra with infrared cutoffs at low wavenumber arising from physics
in the initial phase of a (limited) inflationary epoch.

\begin{figure}
\includegraphics[width=0.85\textwidth]{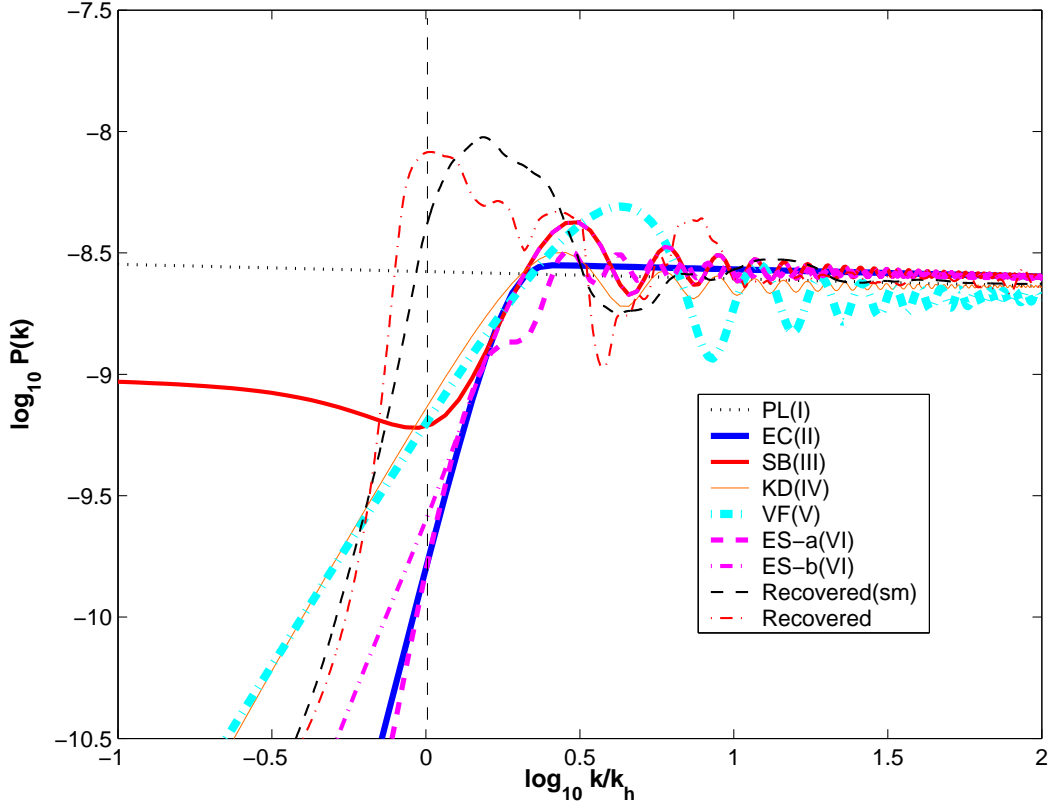} 
\caption{ The class of primordial power spectra with features which
have been used in this work are plotted. Here $k_h=2\pi/\eta_o\approx
4.5 \times10^{-4}$ Mpc$^{-1}$, is the wavenumber corresponding to the
Horizon scale for best fit $\Lambda$CDM model. For comparison we also
give the power spectrum recovered from WMAP data by direct
deconvolution~\cite{arman04} (`Recovered') and wavelet smoothed version that retains
the most prominent features (`Recovered(sm)') (adapted from R. P. Manimaran {\it et
al.}, {\it in preparation}). Notably, none of the model spectra are
able to match up well to the form of infrared cutoff indicated by the
direct deconvolution leaving room for further search of more likely
inflationary scenario.}
\label{pkplot}
\end{figure}

In this section we describe such primordial power spectra with
features which have been considered in this work and each model is
motivated by some plausible inflationary scenario. Fig.~\ref{pkplot}
shows the best fit spectra for the class of primordial power spectra
considered below.

\subsection{Scale-free Power Law model-PL(I)}
Before we go on to describe initial power spectra with features we
would like to mention the commonly assumed power law power
spectrum (referred to as `Power Law' model here). This is generally parametrized in terms of the slow-roll
approximation of the single-inflaton field and is given by
\begin{equation}
\mathcal{P}_o(k) = A_s \, {\left( {k\over k_0} \right)}^{n_s-1} \,.
\label{plspec}
\end{equation}
Here $n_s(k) = [ d \ln \mathcal{P}(k)/d ln k ] + 1$ is the scalar
spectral index which is constant for power law models ($n_s = 1$
corresponds to a scale-invariant power spectrum) and $k_0$ is the
scalar pivot scale. The power spectrum amplitude, $A_s$ determines the
variance of the fluctuations.

The various inflation motivated primordial power spectra
models~\cite{fnote4} with infrared cutoffs (with or without a `bump')
which have been explored in this work are described below.

\subsection{Exponential Cutoff model-EC(II)}
This model has a monotonic cutoff imposed on (multiplying) a
scale-free power law spectrum. Such an infrared cutoff discussed and
motivated by~\cite{contaldi03_supcmb} leads to power spectrum of the form (referred to as `Expo-cutoff' model here)
\begin{equation}
\mathcal{P}(k)\,=\,A_s\,k^{n_s-1}\,\left[1-e^{-(k/k_*)^{\alpha}}\right]  \,.
\end{equation}
In previous analyses using this model in \cite{{contaldi03_supcmb},{cline03}}
the value of the parameter $\alpha$ was fixed ($\alpha= 3.35$) to
approximately mimic the sharpness of the cutoff in a specific scenario
(See Kin.Dom model described later). In our work here we allow it to vary and obtain
the best fit value. As discussed later, our results indicate that a larger value of
$\alpha$ is preferred.

\subsection{Starobinksy model-SB(III)}
It was shown by Starobinsky~\cite{star92} that if the effective scalar
inflaton potential has a `singularity' in the form of a sharp change
in the slope (a `kink'), it can create an infrared suppression in power
spectrum of adiabatic perturbations at any chosen wavenumber $k_*$ (referred to as 'Starobinsky' model here). The infrared cutoff is followed by the `bump' that arises naturally as the first peak of a damped
ringing.  The effect of a kink (sharp change in the slope) at some
point of the inflaton potential can be neatly expressed in terms of the
analytic multiplicative transfer function applied on the underlying
power spectrum $\mathcal{P}_o(k)$ as

\begin{equation}
\mathcal{P}(k) \, =\, \mathcal{P}_o(k) \, \,  \mathcal{D}^2(y,R_{*}) \, .
\label{staropk}
\end{equation}
In the simpler form (SB) we restrict $\mathcal{P}_o(k)$ to be a simple power law. However, in general, $\mathcal{P}_o(k)$ can be of any form allowed by models of scalar field driven inflation. For example, in the later section we take it to be of an exponential form (Expo-Staro model). The transfer function
\begin{eqnarray}
\mathcal{D}^2(y,R_{*}) \, = \, [1-3\,(R_{*}-1)\, \frac{1}{y}\, ((1-\frac{1}{y^2})\sin2y+\frac{2}{y}\cos2y)  
\,+\, \frac{9}{2}(R_{*} -1)^2\frac{1}{y^2}(1+ \frac{1}{y^2}) \, \times \nonumber\\ 
(1+\frac{1}{y^2} +(1-\frac{1}{y^2})  \cos2y-\frac{2}{y}\sin2y) ] \, ,
\label{staro}
\end{eqnarray} 
with $y={k}/{k_*}$. $R_{*}$ is the ratio of the slope $dV/d\phi$ which
dictates the shape of the power spectrum, $V(\phi)$ being the inflaton
potential of the scalar field $\phi$. The power spectrum
$\mathcal{P}(k)$ in eq.~\ref{staropk} has a step-up like feature (going to
larger $k$) for $R_{*}<1$ and a step-down like feature for $R_{*}>1$.
An infrared cutoff is created when $R_{*}<1$. This shape of the
primordial power spectrum with a Starobinsky step not only mimics the
sharp infrared cutoff but also subsequently produces the required
localized excess in power (the `bump'). We have carried out MCMC
analysis and obtained best fit values for $k_*$ and $R_{*}$.

\subsection{Pre-inflationary Kinetic Domination model-KD(IV)}
The observable inflationary epoch could be preceded by a period of
fast rolling of the inflaton field $\phi$ leading to pre-inflationary
phase of kinetic domination. The difference of the vacuum in the
kinetic domination (fast rolling) phase relative to the inflationary phase would imprint a feature in the
power spectrum at large scales corresponding to first modes that
crossed out of the Hubble radius at the onset of
inflation~\cite{contaldi03_supcmb}. The feature is an infrared cutoff akin to
that first shown by Vilenkin and Ford ~\cite{vilkford82} for a
radiation dominated pre-inflationary phase.  If the scale corresponding
to the current horizon exited the horizon very soon after the onset of
inflation, then the feature could explain the observed suppression of
power at the horizon scale. The form of primordial perturbations is
given by (referred to as `Kin. Dom.' model here)
\begin{equation}
\mathcal{P}(k) \, =\, {{H^2}\over{2\pi^2}} \, k \, {\mid A -  B \mid}^2 \,
\end{equation} 
where
\begin{eqnarray}
A \, = \, \frac{e^{-ik/H}}{\sqrt{32 \, H/\pi}} \left[ {\mathcal{H}}_0^{(2)} \left(
\frac{k}{2H} \right) \, - \, \left( \frac{H}{k}\, + \, i \right) \,
{\mathcal{H}}_1^{(2)} \left( \frac{k}{2H} \right) \right] \nonumber
\end{eqnarray} 
and
\begin{eqnarray}
B \, = \, \frac{e^{ik/H}}{\sqrt{32 \, H/\pi}} \left[ {\mathcal{H}}_0^{(2)} \left(
\frac{k}{2H} \right) \, - \, \left( \frac{H}{k}\, - \, i \right) \,
{\mathcal{H}}_1^{(2)} \left( \frac{k}{2H} \right) \right] \, .  \nonumber
\end{eqnarray} 
Here $H$ denotes the (physical) Hubble parameter during inflation
while $\mathcal{H}_0^{(2)}$ and $\mathcal{H}_1^{(2)}$ denote the Hankel function of the
second kind with order 0 and 1, respectively (for details refer
to~\cite{contaldi03_supcmb} and references therein).

\subsection{Pre-inflationary Radiation domination model-VF(V)}
For a pre-inflationary radiation dominated epoch the power spectrum was given 
by Vilenkin and Ford ~\cite{vilkford82} as follows (referred to as `VF' model here) 

\begin{equation}
P(k) = A_s\,k^{1-n_s}\,\frac{1}{4 y^4}\left| e^{-2 i y} (1+2 i y)
-1 - 2 y^2\right|^2 \, ,
\label{vf}
\end{equation} 
where $y={k}/{k_*}$. The VF power spectrum can also provide an
infrared cutoff with the required `bump' after it.  The infrared
cutoff here is not as sharp as the one arising from kinetic
domination in the pre-inflationary phase and is found to be relatively
disfavored in our work. The cutoff scale $k_*$ is set by the Hubble
parameter at the onset of inflation. Here too, the current horizon
scale crosses the Hubble radius very close to the onset of inflation.

\subsection{Exponential-Starobinsky model-ES(VI)}
Encouraged by the reasonable success of the class of primordial
spectra described in the earlier subsections we have undertaken a power
estimation with yet another form of initial spectra tailored to mimic
the cutoff and the `bump'  in the primordial power spectrum.  As
discussed earlier, the Starobinsky feature is a transfer function that
modulates an underlying power spectrum expected from the inflaton potential without singularity and affects any inflationary scenario based on a inflaton potential with sharp change in slope (eq.~\ref{staro}). We
imprint a Starobinsky break on an exponential cutoff spectrum (referred to as `Expo-staro' model here) mainly motivated by the fact that these models provide a sharp reduction in large wavelength power needed to suppress the low multipoles of the CMB.  
The Expo-staro model has both the exponential cutoff providing an adjustable sharpness
of the cutoff as well the bump of the Starobinsky feature and is more fine tuned as reflected by the additional
parameter, $\varepsilon_*$. This can be described as 
\begin{equation}
\mathcal{P}(k) \, =\, \, \mathcal{P}_o(k)\, \left[1-e^{-({\varepsilon_*} \, {k/k_*})^{\alpha}}\right] \, \,  \mathcal{D}^2(y,R_{*}) \, ,
\end{equation} 
where $\mathcal{D}^2(y,R_{*})$ is the transfer function of the
Starobinsky feature described by eq.(~\ref{staro}) and $\varepsilon_*$
sets the ratio of the two cutoff scales involved (exponential cutoff
and Starobinsky feature).

\bigskip
Using the recent measurement of the angular power spectrum of CMB
anisotropy we undertake a full blown multi-parameter joint estimation
of the parameters characterizing the infrared cutoff such as $k_*$,
$\alpha$ and $R_{*}$ (described above) together with the standard set
of cosmological parameters. In the next section we briefly describe the
parameter estimation method used in this work.

\section{Methodology}

Theoretical models for the determination of the power spectrum, which
is responsible for structure formation, starts from the assumption of
a primordial $\mathcal{P}_{o}(k)$ at a sufficiently high redshift $z \gg
z_{eq}$, the redshift at matter-radiation equality. The primordial power spectrum is 
related to the linear CMB anisotropy
power spectrum $C_l^{M_{12}}$ by the transfer function
$T_l^{M_{12}}(k)$ such that~\cite{bridle03}
\begin{equation}
C_l^{M_{12}} \, \propto \, \int \, {\rm d} \, {\rm ln} k \, \mathcal{P}(k) \, T_l^{M_1} (k) \, T_l^{M_2} (k)
\end{equation}
where $M_1$ and $M_2$ are the indices for the temperature and the two polarization
modes of the power spectra.  

We use the well-known {\it Monte Carlo Markov Chain} (MCMC) technique
~\cite{{radfordMC},{cosmo}} to constrain the parameters with the
observed data. Given a vector of cosmological data {\bf d} (for
instance, {\bf d} can be the vector containing the 899 WMAP
measurements of the temperature power spectrum $C_l$ for $l=2$ to 900)
from which we want to measure a vector of cosmological parameters {\bf
p} $\equiv$ ($\tau, \Omega_m, \Omega_b, \Omega_{\Lambda}$,
$k_*$,...). The cosmological parameterization has been carried out in
terms of the following parameters, namely, the baryon density
$\Omega_b{h^2}$, cold dark matter density, $\Omega_{cdm}h^2$, Hubble
constant $H_o$~\cite{fnote6}, and opacity to
the surface of last scattering, $\tau$~\cite{fnote7}. Apart from these, we include
the parameters characterizing the inflationary primordial
perturbations with infrared cutoff features as described in the
previous section ($k_{*}$, $\alpha$, $R_*$) along with the scalar
spectral index of perturbations, $n_s$~\cite{fnote8} and the amplitude
log[$10^{10}A_s$]. We consider a flat universe with $\Omega_{k}=0$. Table~\ref{paramprior} gives the priors used in the study for the various inflationary and cosmological parameters. The ranges of these uniform priors have been varied to ensure that there is no dependence of priors on our results.  We restrict our study to a cosmological flat model where the cosmological constant, $\Omega_{\Lambda}=1- \Omega_{cdm} - \Omega_b$ and do not consider other forms of dark energy. The theory {\bf p} is connected to the data {\bf d} by the likelihood
function, $\mathcal{L}$({\bf p}, {\bf d}), which gives the probability
distribution for observing different {\bf d} given a theoretical model
{\bf p} which in our case is the $\Lambda$CDM model. In Bayesian
analysis, one inserts the actual observed data and and reinterprets
$\mathcal{L}$({\bf p}, {\bf d}) as an un-normalized probability
distribution over the cosmological parameters {\bf p}, optionally
after multiplication by a probability distribution, which has the prior
information. To place constraints on any single parameter, one
marginalizes (integrates) over all the others.  In the MCMC method a
large set of points {\bf p}$_i, i=1$, \dots n, i.e. a {\it Markov}
chain, is generated by a stochastic procedure such that the points
have a probability distribution $\mathcal{L}$({\bf p}, {\bf d}).  The
basic MCMC algorithms used here are based on the Metropolis-Hastings
and slice sampling methods. According to Metropolis-Hastings
rule~\cite{metrop53} higher likelihood trials are always accepted
i.e. for $\mathcal{L}$({\bf p$_*$}, {\bf d}) $>$ $\mathcal{L}$({\bf
p$_i$}, {\bf d}), otherwise rejected and accepted only with
probability $\mathcal{L}$({\bf p$_*$}, {\bf d})/$\mathcal{L}$({\bf
p$_i$}, {\bf d}) where {\bf p}$_*$ is a new trial random point in the
parameter space.

The other method used in the implementation of MCMC is slice sampling~\cite{cosmo}. In slice sampling~\cite{slice}
similar to the Metropolis method the target density {\bf p}$_*(x$) is
evaluated at any point $x$.  Here is a brief note on slice sampling
algorithm. Consider a one-dimensional slice sampling method for making
transitions from a two-dimensional point $(x, u)$ lying under the
curve {\bf p}$(x)$ to another point ($x', u')$ lying under the
same curve, such that the probability distribution of $(x, u)$ tends
to a uniform distribution over the area under the curve {\bf
p}$_*(x)$, whatever initial point we start from.  A single transition
$(x, u) \rightarrow (x', u')$ of a one-dimensional slice sampling
algorithm has the following basic steps. A vertical coordinate $u'$
and an horizontal interval $(x_a,x_b)$ enclosing $x$ (i.e. a slice) is
then drawn under the curve {\bf p}$_*(x)$. Now $x'$ is drawn between
this interval and {\bf p}$_*(x')$ is evaluated. If {\bf p}$_*(x') >
u'$ then accept it otherwise choose a different interval and
continue. There can be slight variations to this, for instance, `stepping
out' or `shrinking' methods can be used for choosing the interval $(x_a,x_b)$, the
distribution under the curve always remaining uniform.  Slice sampling
is more robust to the choice of parameters like step sizes which is
self tuning in this method and hence, we get better estimates
when implementing MCMC with the new parameter set. However, slice
sampling is slightly slower in terms of computer run-time.

We have evaluated the likelihood function $\mathcal{L}$ and obtained
the best fit from the parameter searches using a modified version of
CosmoMC~\cite{cosmo}. We typically ran multiple chains on a 64 bit 8x4
Alpha-Server ES45 68/1250 System (1.25 GHz each) to optimize our
parameter set.  We have first conducted the analysis with only WMAP TT
data wherein the discrepancy in power is more apparent. Later on the
analysis has been extended to include both the WMAP TT and TE angular
power spectra along with the CBI (Cosmic Background Imager),
ACBAR (Arcminute Cosmology Bolometer Array Receiver) and VSA (Very
Small Array) datasets of the CMB for small scales ($l\gtrsim
700$). These external datasets augment the quality of the analysis as
well as help in alleviating parameter degeneracies. For instance, these are essential to break parameter degeneracy between the inflationary parameter, $n_s$ and optical depth, $\tau$. In addition, high resolution CMB datasets also help to tightly constrain the parameters for example, $\Omega_m$, amplitude $A_s$, the running in the spectral index (not used here) ~\cite{wmap_pest, contaldi03_jntcmb, wmap3y_spergel06}.

\begin{table} 
\caption{Uniform prior ranges for the parameters as used in the joint parameter estimation study.}
\begin{center}
\begin{ruledtabular}
\begin{tabular}{lccc}
& Parameter & Prior (range) \\ 
\hline
\\
Baryon Density & $\Omega_b{h}^2$ & 0.005-0.1 & \\ \\
Dark Matter Density & $\Omega_{dm}{h}^2$ & 0.01-0.99 & \\ \\
Angular size of Acoustic Horizon\footnote[1]{Acoustic horizon as defined in ~\cite{fnote6}} & $\theta$ & 0.5-10.0 & \\ \\
Optical Depth & $\tau$ & 0.01-0.8 & \\ \\
Scalar Spectral Index & $n_s$  &  0.5-1.5 & \\ \\
Infrared Cutoff (wavenumber) & $k_{*}$($\times 10^{-4}$)Mpc$^{-1}$ & 0.00001-0.0009 & \\ \\
Expo Cutoff Steepness Parameter & $\alpha$ & 0.1-10.0 & \\ \\
Starobinsky Parameter & $R_{*}$ & 0.01-0.99 & \\ \\
Amplitude Parameter\footnote[2]{Amplitude parameter as defined in ~\cite{fnote61}} & log$10^{10} A_s$ & 2.7-4.0 & \\
\label{paramprior}
\end{tabular}
\end{ruledtabular}
\end{center}
\end{table}

\section{Parameter Estimates}

We have carried out a complete Bayesian analysis by varying and marginalizing 
over all the parameters considered in this work to obtain the best fit values. 
Table~\ref{paramsbw} and Table~\ref{paramsbwe} sums the results of the MCMC simulations with the WMAP data only and WMAP data plus the addition of the external datasets, respectively. Tables II and III mention the values of some of the relevant parameters (the search has been made for a larger parameter space) along with their 1-$\sigma$ error bars and their goodness-of-fit parameter ($\chi^2_{eff} \equiv- 2 \, \ln \mathcal{L}$) values for the different models.

\begin{table} 
\caption{Best fit values of parameters specifying the initial power
spectrum ($k_{*}, \alpha,R_{*}, n_s$) and other relevant cosmological
parameters for a class of model power spectra with a infrared cutoff (dataset
used: WMAP TT data).}
\begin{center}
\begin{ruledtabular}
\begin{tabular}{lcccccccc}
Parameter & Expo-cutoff & Starobinsky & Kin. Dom. & VF & {\bf{Expo-staro(a)}}$^{\dagger}$\footnote[0]{$^\dagger\varepsilon_*=0.2$} & Expo-staro(b){$^\ddagger$}\footnote[0]{$^\ddagger\varepsilon_*=0.75$}   & Power Law \\
& EC(II) & SB(III) & KD(IV) & VF(V) & ES-a(VI) & ES-b(VI) & PL(I) \\ 
\hline
\\
$k_{*}$($\times 10^{-4}$)Mpc$^{-1}$ & 3.0$_{-2.9}^{+4.8}$ & 3.1$_{-2.8}^{+5.8}$ & 3.5$_{-3.3}^{+3.0}$ &  0.4$_{-0.3}^{+0.7}$ & {\bf 3.0${\bf_{-2.0}^{+0.5}}$}  & 3.1$_{-2.1}^{+5.8}$ & $-$ & \\ \\
$\alpha$ & 9.6$_{-8.6}^{+0.3}$ &$-$ & $-$ & $-$ & {\bf 0.58${\bf_{-0.43}^{+4.6}}$} & 0.72$_{-0.55}^{+9.1}$ & $-$ & \\ \\
$R_{*}$ & $-$ & 0.73$_{-0.14}^{+0.25}$ & $-$ & $-$ & {\bf 0.17${\bf_{-0.15}^{+0.80}}$} &  0.35$_{-0.20}^{+0.63}$ & $-$ & \\ \\
$n_s$  &  0.95$_{-0.03}^{+0.16}$ & 0.98$_{-0.07}^{+0.14}$ & 1.4$_{-0.90}^{+0.09}$ & 1.0$_{-0.15}^{+0.04}$ & {\bf 0.96${\bf_{-0.08}^{+0.15}}$} &  0.99$_{-0.12}^{+0.08}$ & 0.96$_{-0.05}^{+0.30}$ & \\ \\
$\tau$ & 0.014$_{-0.004}^{+0.37}$  & 0.15$_{-0.14}^{+0.25}$  & 0.17$_{-0.15}^{+0.09}$ & 0.01 $_{-0.001}^{+0.35}$ & \bf {0.26${\bf_{-0.08}^{+0.15}}$} &  0.28$_{-0.27}^{+0.12}$ & 0.014$_{-0.004}^{+0.500}$ & \\ \\
$z_{re}$\footnote[1]{Reionization epoch as defined in ~\cite{fnote7}} & 3.2$_{-0.7}^{+21.7}$  & 16.3$_{-13.9}^{+11.5}$ & 17.8$_{-15.2}^{+4.9}$& 2.7$_{-0.22}^{+23.5}$ & {\bf 23.8${\bf_{-5.0}^{+5.9}}$} &  23.5$_{-21.0}^{+3.9}$ & 3.2$_{-0.83}^{+26.6}$ & \\ \\
$\Omega_\Lambda$& 0.70$_{-0.18}^{+0.16}$ & 0.71$_{+0.24}^{+0.17}$ & 0.70$_{-0.21}^{+0.13}$ & 0.71$_{-0.20}^{+0.12}$ & {\bf 0.74${\bf_{-0.10}^{+0.13}}$} &  0.75$_{-0.23}^{+0.12}$ & 0.65$_{-0.23}^{+0.24}$ & \\ \\
$\Omega_b{h}^2$& 0.022$_{-0.001}^{+0.006}$ & 0.023$_{-0.004}^{+0.005}$ & 0.024$_{-0.002}^{+0.001}$ & 0.023$_{-0.002}^{+0.005}$ & {\bf 0.023${\bf_{-0.003}^{+0.004}}$} & 0.025$_{-0.005}^{+0.002}$ & 0.023$_{-0.002}^{+0.009}$ & \\ \\
-$\ln{\cal{L}}$ & 484.89 & 484.89 & 485.18 & 486.46 & {\bf 483.44} & 484.45 & 486.28 & \\ \\
${\chi_{\rm eff}^2} \equiv -2 \ln{\cal{L}}$ & 969.78 & 969.78 & 970.36 & 972.92 & {\bf 966.88} & 968.90 & 972.56 & \\ \\
d.o.f. & 891 & 891 & 892 & 892 & {\bf 890} & 890 & 893 & \\
\label{paramsbw}
\end{tabular}
\end{ruledtabular}
\end{center}
\end{table}

\begin{figure}
\includegraphics[angle=270,width=0.9\textwidth]{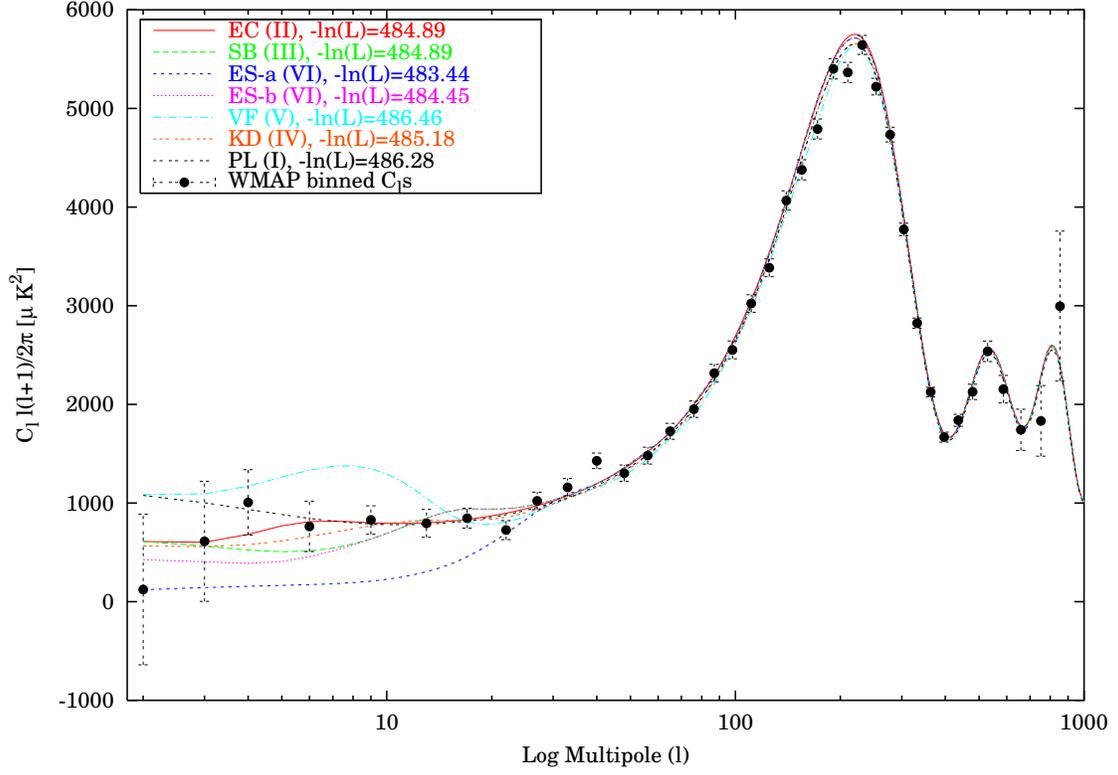}
\caption{ The angular power spectra ($C_l$) for the different
primordial power models along with their best fit likelihood values are given in the figure. 
The plot also shows the WMAP binned C{$_l$}s for comparison}.
\label{clplot}
\end{figure}

The best likelihood value (with the WMAP TT data) is given by
ESa-b(VI) whereas EC(II) and SB(III) likelihood values are
comparable. We obtain $\varepsilon_*=0.2$ (ES-a(VI)) by carrying out
MCMC with ES(VI) model which gives the best likelihood for this model
while $\varepsilon_*=0.75$ (ES-b (VI)) gives the best shape (i.e. closest to the recovered power spectrum)  but is slightly less likely.  In Fig.~\ref{clplot} we have
plotted the angular power spectrum $C_l$ obtained using the best fit
values after the MCMC analysis.  We corroborate the preferred value
for $k_*\sim 3 \times 10^{-4}$ Mpc$^{-1}$ as obtained
by~\cite{cline03} for the exponential cutoff power spectrum (EC(II)). An important
difference to be noted is in the value of $\alpha$.  We obtain better
likelihoods as the value of $\alpha$ increases. In fact, it can be
seen from Table~\ref{paramsbw} that the best likelihood is obtained for a steeper
exponential cutoff, i.e. $\alpha=9.6$ ($-\ln \mathcal{L}=485.2$ for
$\alpha=3.35$) in contrast to earlier studies~\cite{cline03}. This is
also in good agreement with the recovered power spectrum of
~\cite{arman04} where again a very sharp cutoff ($\alpha=10$), gave a
better fit to WMAP data.
 
\begin{figure}
\includegraphics[angle=270, width=0.85\textwidth]{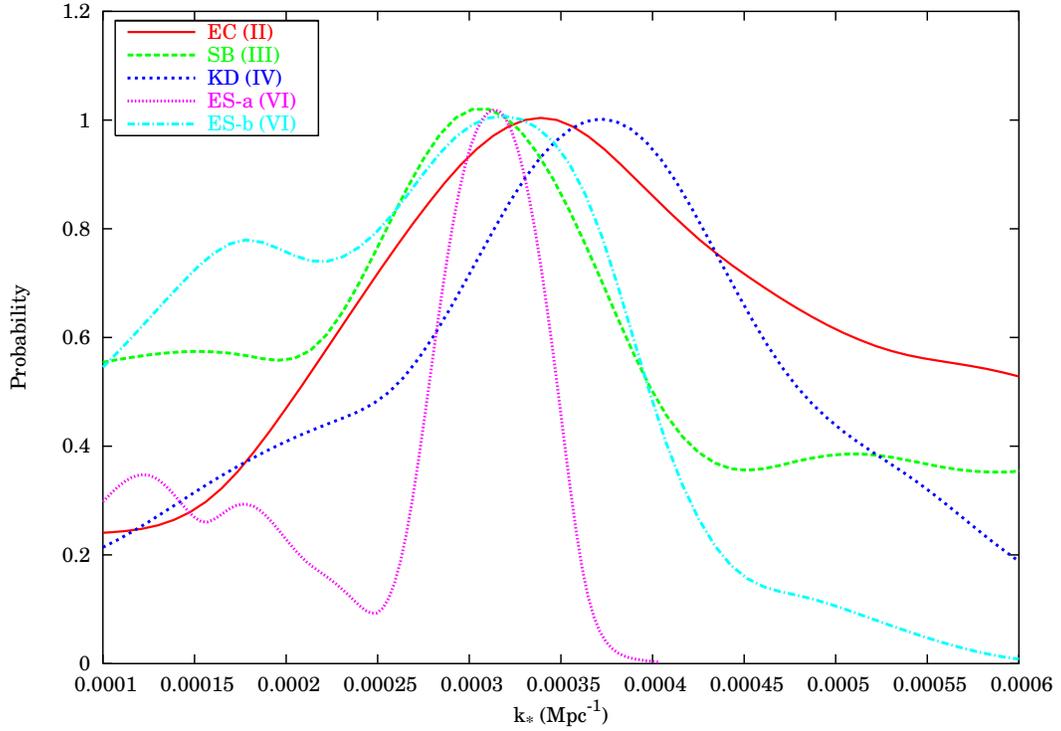}
\caption{The plot shows the mean likelihoods of the cutoff parameter,
$k_*$ obtained after performing MCMC for the models as discussed in
the text using WMAP TT data. Notably, the probability for model
ES-a(VI) is sharply peaked around $k_*\sim 3.2\times10^{-4}$ Mpc$^{-1}$.}
\label{probkcut}
\end{figure}

Fig.~\ref{probkcut} shows the mean likelihoods for $k_*$ for the
models using WMAP TT data. We note that compared to other models
the probability for model ES-a(VI) is sharply peaked on $k_*\sim
3.2\times10^{-4}$ Mpc$^{-1}$. This however, does not translate to
tight 1-$\sigma$ error bars in the corresponding table due to the
bimodal probability distribution with a smaller, broad secondary peak
at low wavenumber for this model.

\begin{figure}
\centering
\framebox{
\includegraphics[width=0.37\textwidth]{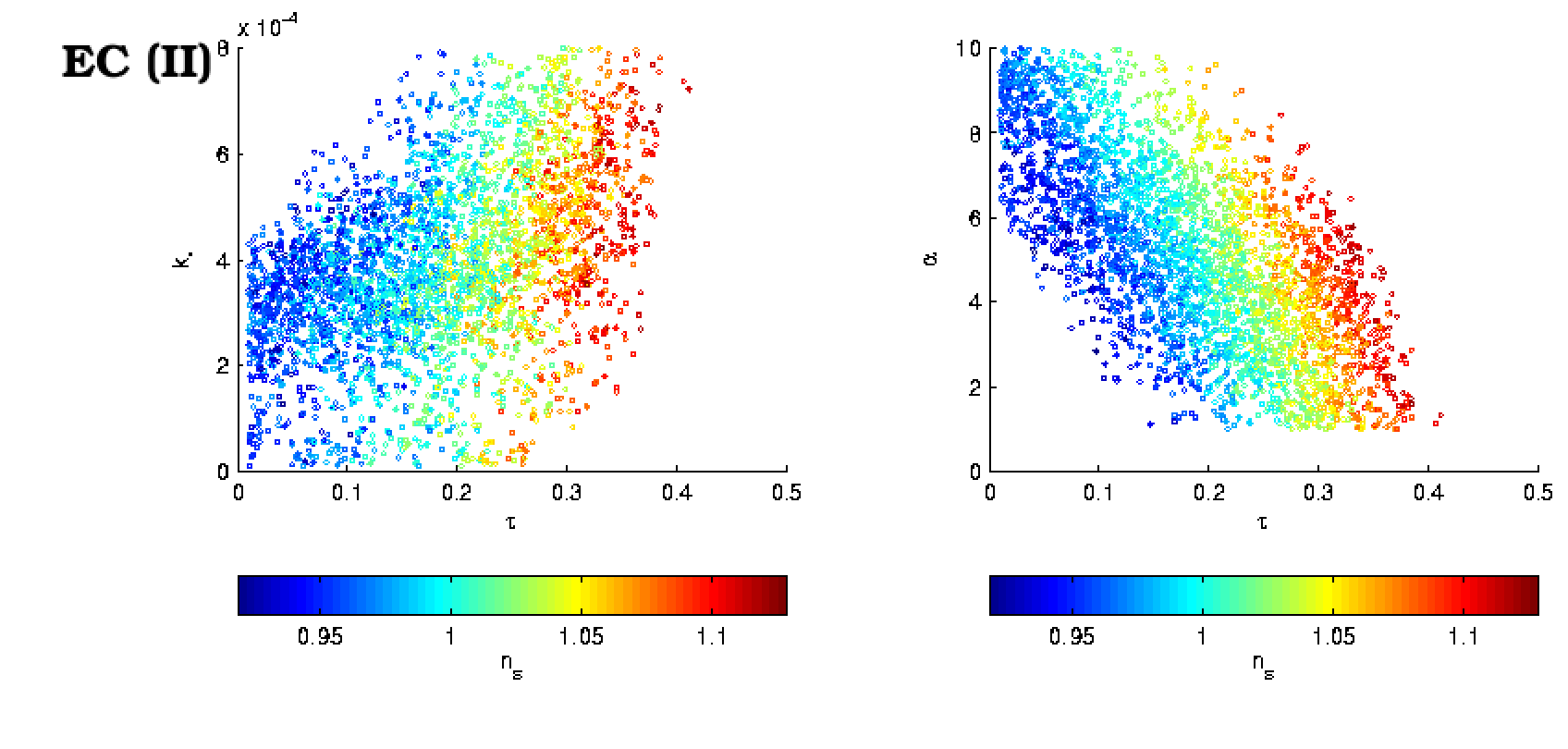}}
\framebox{
\includegraphics[width=0.40\textwidth]{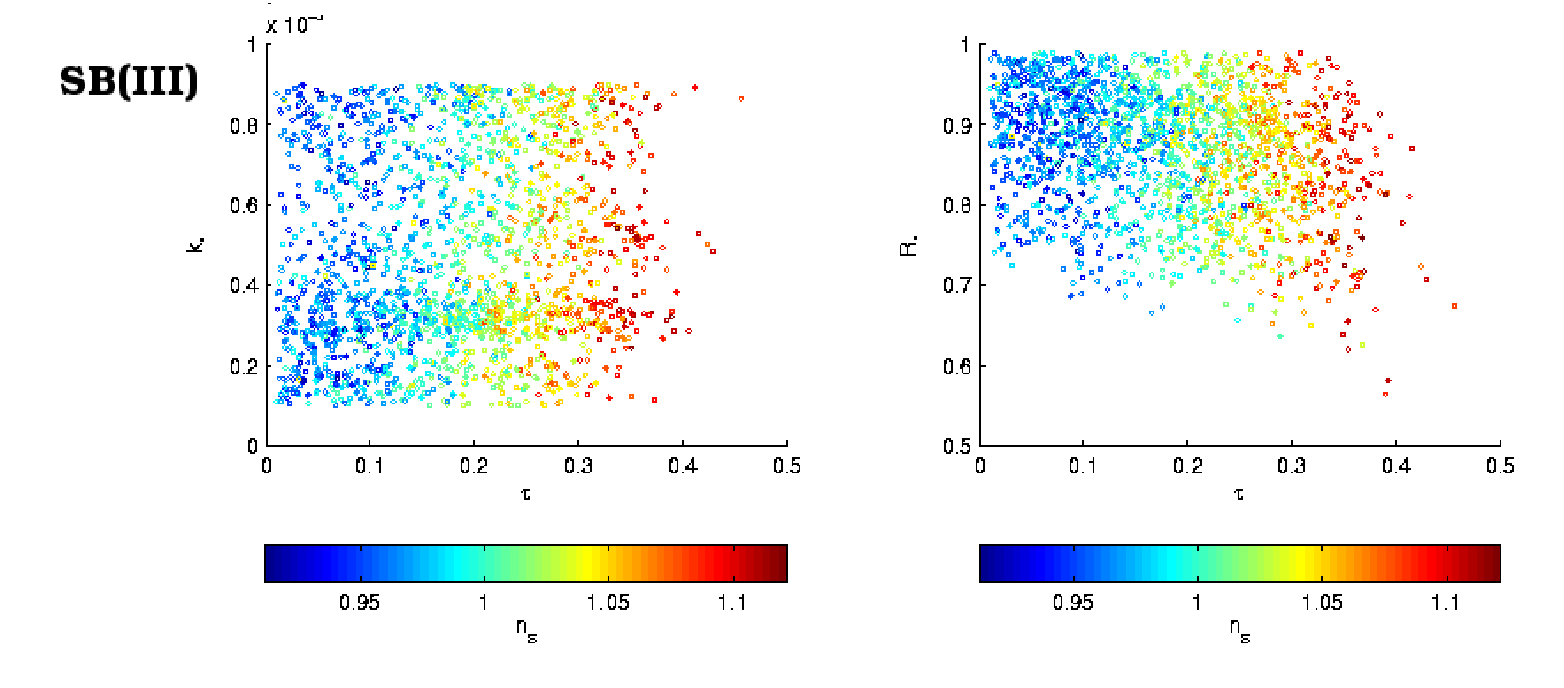}}
\\
\framebox{
\includegraphics[width=0.7\textwidth]{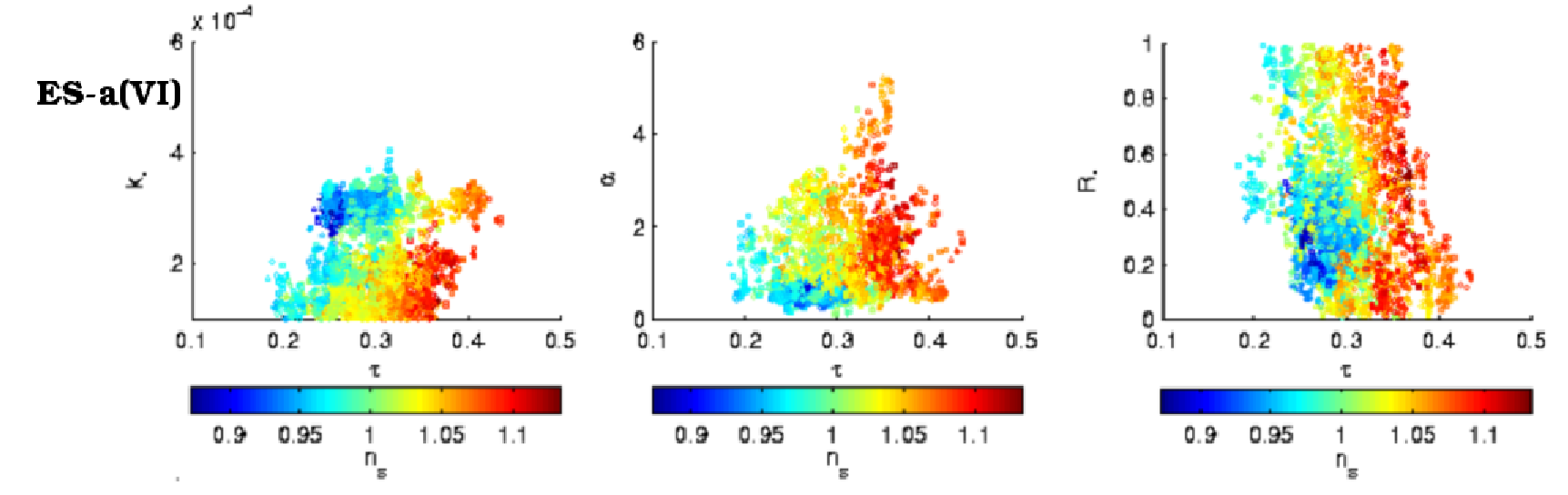}}
\caption{The plots show the posterior distribution of the parameter
$\tau$ plotted against the parameters $k_{*}$ and $\alpha$ for
exponential cutoff EC(II) model (Top left panel), $k_{*}$ and $R_{*}$ for
Starobinksy SB(III) model (Top right panel) and $k_*$, $\alpha$ and
$R_{*}$ for expo-staro ES-a(VI) model (bottom panel). Points are
colored according to the value of $n_s$.}
\label{pdist}
\end{figure}

The values of the spectral index, $n_s$ and cosmological constant,
$\Omega_{\Lambda}$ affect the shape and the amplitude of the power
spectrum of matter distribution.  We would like to mention that
getting the estimate of $\tau$ from the low multipoles is important to
break parameter degeneracies. As more parameters are added,
precision continues to reduce (unless new degeneracies are broken) as
many of the parameters are correlated. Deviation in primordial spectral
index $n_s$ is a consequence of strong correlation with
$\tau$~\cite{bernd01}. In Fig.~\ref{pdist} we show the correlation
between the optical depth $\tau$ and the inflationary parameters,
$k_{*}$, $\alpha$, $R_{*}$ along with the correlated $n_s$. We notice
that the increase in $\tau$ and $k_*$ is compensated by a corresponding increase in
$n_s$. Therefore, we find that inflationary models such as SB(III),
KD(IV), and ESa-b(VI) models (with cutoff plus `bump') provide
better fit to the observed WMAP data. Another interesting
result that emerges is the difference in $\tau$ inferred with and
without the WMAP polarization TE data. It is much less marked for
models with infrared cutoff than for the pure Power Law model. Perhaps
this could be an indirect evidence in favour of an infrared cutoff.
We consider the VF model as not a very good fit
to the data and leave it out of any further discussion.  Table~\ref{paramsbwe}
gives the parameter search results with the external CMB datasets
(CBI, ACBAR, VSA). In this case too we report the best likelihood with
ESa-b(VI) $\mathcal{P}(k)$ compared to the rest of the pack. To investigate the effect of priors we performed a 
number of MCMC simulations with different priors within the ranges given in Table~\ref{paramprior} and found that the 
dependence of the results on priors is weak ~\cite{slosar03}.

\begin{table} 
\caption{Best fit values of parameters specifying the initial power
spectrum ($k_{*}, \alpha,R_{*}, n_s$) and other relevant cosmological
parameters obtained using a class of model power spectra with a infrared cutoff
(dataset used: WMAP (TT+TE) + External (CBI, ACBAR, VSA) CMB
datasets)}
\begin{center}
\begin{ruledtabular}
\begin{tabular}{lcccccccc}
Parameter & Expo-cutoff & Starobinsky & Kin. Dom. & VF & {\bf Expo-staro(a)}$^{\dagger}${\footnote[0]{$^\dagger\varepsilon_*=0.2$}} & Expo-staro(b){$^\ddagger$}\footnote[0]{$^\ddagger\varepsilon_*=0.75$} & Power Law \\
& EC(II) & SB(III) & KD(IV) & VF(V) & ES-a(VI) & ES-b(VI) & PL(I) \\ 
\hline 
\\
$k_{*}$($\times 10^{-4}$) Mpc$^{-1}$ & 2.6$_{-2.5}^{+5.2}$ & 8.9$_{-7.9}^{+0.1}$ & 3.5$_{-3.4}^{+2.9}$ &   0.40$_{-0.30}^{+1.7}$ & {\bf 1.2${\bf_{-0.19}^{+1.0}}$} & 1.3$_{-0.3}^{+2.5}$ & $-$ & \\ \\
$\alpha$ & 6.0$_{-4.9}^{+3.5}$ & $-$ & $-$ & $-$ &  {\bf 0.89${\bf_{-0.73}^{+0.41}}$} &  0.78$_{-0.64}^{+2.4}$ & $-$ & \\ \\
$R_{*}$ & $-$ & 0.85$_{-0.5}^{+0.14}$ & $-$ & $-$ & {\bf 0.017${\bf_{-0.006}^{+0.96}}$} & 0.15$_{-0.14}^{+0.83}$ & $-$ & \\ \\
$n_s$  &  0.96$_{-0.04}^{+0.09}$ & 0.95$_{-0.06}^{+0.16}$ & 1.40$_{-0.97}^{+0.02}$ & 1.04$_{-0.11}^{+0.05}$ & {\bf 0.95${\bf_{-0.09}^{+0.11}}$} & 0.94$_{-0.07}^{+0.15}$ & 0.95$_{-0.05}^{+0.19}$ & \\ \\
$\tau$ & 0.12$_{-0.10}^{+0.20}$  & 0.19$_{-0.18}^{+0.23}$  & 0.23$_{-0.11}^{+0.04}$  & 0.12$_{-0.07}^{+0.23}$ & {\bf 0.16${\bf_{-0.09}^{+0.21}}$} & 0.12$_{-0.31}^{+0.36}$ &  0.11$_{-0.09}^{+0.35}$ & \\ \\
$z_{re}$\footnote{Reionization epoch as defined in ~\cite{fnote7}} & 13.8$_{-10.6}^{+11.3}$  & 19.1$_{-16.4}^{+9.05}$ & 20.7$_{-7.2}^{+1.7}$ & 14.2$_{-6.1}^{+11.1}$ & {\bf 16.9${\bf_{-7.6}^{+10.7}}$} & 13.6$_{-6.5}^{+16.6}$ & 13.1$_{-10.4}^{+14.9}$ & \\ \\
$\Omega_\Lambda$& 0.74$_{-0.10}^{+0.12}$ & 0.78$_{-0.15}^{+0.12}$ & 0.80$_{-0.12}^{+0.06}$ & 0.73$_{-0.10}^{+0.13}$ & {\bf 0.75${\bf_{-0.16}^{+0.10}}$} & 0.72$_{-0.11}^{+0.16}$ & 0.74$_{-0.12}^{+0.15}$ & \\ \\
$\Omega_b{h}^2$& 0.023$_{-0.002}^{+0.003}$ & 0.022$_{-0.003}^{+0.006}$ & 0.024$_{-0.002}^{+0.001}$ & 0.022$_{-0.002}^{+0.004}$ & {\bf 0.022${\bf_{-0.002}^{+0.004}}$} & 0.022$_{-0.003}^{+0.005}$ &  0.023$_{-0.002}^{0.005}$ & \\ \\
-$\ln{\cal{L}}$ & 725.25 & 723.95 & 726.68 & 726.30 & {\bf 723.36} & 723.58 & 726.20 & \\ \\
${\chi_{\rm eff}^2} \equiv -2 \ln{\cal{L}}$ & 1450.50 & 1447.90 & 1453.36 & 1452.60 & {\bf 1446.72} & 1447.16 & 1452.40 & \\ \\
d.o.f. & 1442 & 1442 & 1443 & 1443 & {\bf 1441} & 1441 & 1444 & \\
\label{paramsbwe}
\end{tabular}
\end{ruledtabular}
\end{center}
\end{table}

We note that introducing a cutoff $k_{*}$ and a `bump' at low
wavenumber in the primordial power spectrum provides a better fit to
the observed CMB data.  Such scenarios can explain the lack of power
at low multipoles as shown in Fig.~\ref{clplot}.  

\section{Discussion and Conclusion}

The cosmological model and parameter constraints deduced from
observations are sensitive to assumptions regarding the power spectrum
of primordial density perturbations. The suppression of power in the
quadrupole of the CMB anisotropy could be signaling a radical
departure (in the form of an infrared cutoff) from the near-scale
invariance of the primordial spectrum predicted by simplest
inflationary scenarios. The very presence of a feature in the
primordial spectrum indicates modification to the simplest
inflationary scenarios and calls for an exploration of extensions based
on plausible physical mechanisms.

We show that sharp features, such as a infrared cutoff near the horizon scale
and a `bump', in the primordial power spectrum provides a likely
explanation to power suppression at low multipoles as seen in the
recent WMAP data. Although a steep monotonic cutoff tends to pull down
the power in the next few higher multipoles above the quadrupole and
octopole but this is well compensated by the sudden excess power
following the sharp rise. This allows a steep cutoff to match the low
quadrupole and octopole without suppressing the higher multipoles.
This is evident in the studies where the primordial power spectra was 
directly deconvolved from the data~\cite{arman04,bump04}. In the light
of the analysis conducted in this work we give more credence to
inflationary models with cutoffs than put forth in earlier
studies~\cite{contaldi03_supcmb,cline03}.  It can be seen through this study
that a break in the initial power spectrum confronts the CMB data
better.  This is also borne out in recent Bayesian evidence assessment
of a different set of model spectra with infrared
cutoff~\cite{brid05}.  The question however, remains of the location
of this cutoff. Our study suggests that a cutoff of $k_* \sim 3 \times
10^{-4}$ Mpc$^{-1}$ with WMAP TT data and $1.1 \times 10^{-4}\leq k_*
\leq 9.0 \times 10^{-4}$ Mpc$^{-1}$ (inclusive of error bars) with all
the CMB datasets.  However, more detailed and wider exploration of the 
infrared cutoff mechanism within inflation is required.  A comparison
of the best fit model spectra with the recovered spectra shows that
the form of infrared cutoff has substantial scope for
improvement. This keeps open the pursuit of (physically motivated)
mechanisms within simple inflation that produces the form of infrared
cutoff suggested by the data.

Studies such as~\cite{BSI98} also assert that the presence of such
features if confirmed further by observations would be attributed to
the initial power spectrum and not due to oscillations produced by any
other means as suggested in Ref.~\cite{barandela97}.  Other feasible
possibilities which produce such features in the initial power
spectrum are the multiple-field inflationary models. Some of these
have been tested with the recent observed CMB data such as the double
inflation~\cite{dble_inflat04} scenario. However, such studies
indicate that the present CMB data (WMAP first year data) strongly
favour single-inflaton field models. Further discussion on the classes
of inflationary models supported by observational data can only be
made and asserted after future advancement in measurements, in
particular, those of CMB polarization. The polarization power spectra have different parameter dependence as opposed to the temperature power spectra. This aids in breaking some parameter degeneracies. Moreover, since their source of origin is different and the polarization spectra directly probes the epoch of last scattering so polarization can be a 
better probe of the inflationary epoch.

\begin{acknowledgments}
We acknowledge the use of IUCAA's High Performance computing
facilities. We acknowledge the use of the CosmoMC package~\cite{cosmo} and are grateful to Anthony Lewis and A. Slosar for useful discussions at the cosmocoffee website
(http://cosmocoffee.info). We are thankful to the WMAP team for the
excellent stride in CMB experiments and for making the data publicly
available. 
\end{acknowledgments}


\begin{thebibliography}{}
\bibitem{fnote1} The quadrupole suffers the most due to the lack of power on large scales followed by the octopole to a lesser extent. In this study we have considered the overall discrepancy observed at low multipoles.
\bibitem{fnote2} {Marginalization over systematic effects such as residuals from foreground removal can further reduce the significance of this discrepancy~\cite{slosar04}.}
\bibitem{fnote3} {It is interesting to recall the similarity of the equation for the generation of perturbations at a given wavenumber and harmonic oscillators. Sharp features in the power spectrum created by changes in the `effective mass' of the perturbations are accompanied by overshoots and ringing just as in (under damped) harmonic oscillators.}
\bibitem{fnote5_irl} The standard  Richardson Lucy (RL) method can start fitting noise at large iterations and produce spurious oscillations. ~\cite{arman04} use an improved Richardson Lucy (RL) method wherein a convergence function is used to weigh down the contribution of noise and sample variance along with a smoothening function (details in ~\cite{arman04}). In this work the plot given in Fig. 1 (Recovered (sm)) is a further improvement over ~\cite{arman04} (Fig. 1 (Recovered)) as it is wavelet smoothed. Such procedure removes any spurious oscillations of the standard RL method giving credence to the features detected by the improved method (~\cite{arman04} has been cited in the WMAP 3-year paper ~\cite{wmap3y_hinshaw06}) .
\bibitem{fnote4} {Although these can be treated under the same class of inflationary models but for the sake of clarity we will hereafter refer to these different forms as different models.}
\bibitem{fnote6} {The Hubble parameter is expressed in terms of the angular scale subtended by acoustic horizon at last scattering i.e. $\theta \, (=r_s/D_a$, i.e. the ratio of the sound horizon, $r_s$ to the angular diameter distance to the last scattering surface, $D_a$).}
\bibitem{fnote7} The redshift at reionization $z_{re}$ is not a free parameter here but a parameter derived from $\tau$ related by  $\tau= \sigma_T\int_0^{z_{re}} dz \, n_e (z )/[(1 + z )^2 H (z )]$, where $n_e$ is the number density of electrons and $\sigma_T$ is the Thompson scattering cross-section, assuming rapid reionization ~\cite{cosmo}.
\bibitem{fnote61} $A_s$ is the primordial superhorizon power in the curvature perturbation on 0.05 Mpc$^{-1}$ scales.
\end{thebibliography}
\end{document}